\documentclass[
  journal=jacsat,
  manuscript=article,
  super=false
]{achemso}

\setcitestyle{numbers,open={(},close={)}}
\usepackage[version=3]{mhchem} 

\usepackage{graphicx}
\usepackage{dcolumn}
\usepackage{bm}
\usepackage{comment}

\usepackage{subcaption}
\usepackage{amsmath}

\usepackage{wrapfig}

\usepackage[justification=centering]{caption}

\usepackage[export]{adjustbox}

\author{Nilesh Dalla}
\email{n.dalla@uw.edu.pl}
\affiliation{University of Warsaw, Faculty of Physics, Institute of Experimental Physics, Ludwika Pasteura 5, Warsaw, 02-093, Poland}

\author{Paweł Kulboka}
\affiliation{University of Warsaw, Faculty of Physics, Institute of Experimental Physics, Ludwika Pasteura 5, Warsaw, 02-093, Poland}

\author{Michał Kobecki}
\affiliation{University of Warsaw, Faculty of Physics, Institute of Experimental Physics, Ludwika Pasteura 5, Warsaw, 02-093, Poland}

\author{Jan Misiak}
\affiliation{University of Warsaw, Faculty of Physics, Institute of Experimental Physics, Ludwika Pasteura 5, Warsaw, 02-093, Poland}

\author{Paweł Prystawko}
\affiliation{Institute of High Pressure Physics “Unipress”, Polish Academy of Sciences, Sokołowska 29/37, Warsaw, 01-142, 
Poland}
\author{Tomasz Kazimierczuk}
\affiliation{University of Warsaw, Faculty of Physics, Institute of Experimental Physics, Ludwika Pasteura 5, Warsaw, 02-093, Poland}

\author{Piotr Kossacki}
\affiliation{University of Warsaw, Faculty of Physics, Institute of Experimental Physics, Ludwika Pasteura 5, Warsaw, 02-093, Poland}

\author{Henryk Turski}
\affiliation{Institute of High Pressure Physics “Unipress”, Polish Academy of Sciences, Sokołowska 29/37, Warsaw, 01-142, Poland}

\author{Tomasz Jakubczyk}
\affiliation{University of Warsaw, Faculty of Physics, Institute of Experimental Physics, Ludwika Pasteura 5, Warsaw, 02-093, Poland}
\email{tomasz.jakubczyk@fuw.edu.pl}

\title[An \textsf{achemso} demo]
  {Efficient Quasi-Resonant, Polarization-Selective Excitation of GaN Quantum Emitters}

\begin{document}

\begin{abstract}

Defect centers in GaN emerge as bright sources of single-photons which recently have been demonstrated to optically interface a localized spin. However, the structure and composition of these defects as well as their efficient excitation techniques were not a subject of thorough studies. This work presents evidence that by tuning the excitation laser energy to specific resonance values the excitation efficiency can be enhanced, resulting in relative increase of photoluminescence intensity by up to an order of magnitude. The resonances can be selectively addressed with linearly polarized light, while the emission dipole remains unchanged, enabling polarization-controlled enhancement. These results establish an efficient way of excitation for GaN-based emitters, thereby increasing the generation rate of photons. The data is consistent with excitation via localized vibrational modes associated with point-defect complexes, establishing a practical quasi-resonant route to brighter, polarization-addressable operation of GaN defect emitters and clarifying their energy-level structure.


\end{abstract}


\section{Introduction}
Gallium nitride (GaN) is widely recognized for its diverse applications in high-speed electronics, high breakdown-voltage devices, and blue light-emitting diodes, owing to its wide band gap\,\cite{roccaforte2020introduction}. While structural defects are generally detrimental for high-performance electronic devices, the optical properties of GaN defects have attracted considerable scientific interest. In particular, defect-related single-photon emission in GaN has emerged as a promising platform for quantum photonics applications. GaN hosts single-photon emitters (SPE's) that operate across a broad spectral range, from near-ultraviolet\,\cite{alkauskasFirstPrinciplesCalculationsLuminescence2012} to near-infrared\,\cite{bishopEnhancedLightCollection2022, lecaronAllfibredTelecomTechnology2025}, offering versatility for quantum communication technologies. These defects have been shown to exhibit bright, stable single-photon emission even at room temperature\,\cite{berhaneBrightRoomTemperature2017}. Recent studies also indicate the existence of a spin degree of freedom associated with visible and IR range defects\,\cite{luoRoomTemperatureOptically2024,engRoomTemperatureOpticallyDetected2025}. The recent demonstration of controlled epitaxy of room-temperature quantum emitters in GaN \cite{EggletonControlledEpitaxyRoomTemperature2026} marks an important step toward scalable quantum photonic technologies based on defect engineering. The emission characteristics are strongly influenced by the defect type, spatial distribution, and substrate selection. Although near-infrared defects are particularly attractive due to reduced optical losses\,\cite{ZhouTelecomGaNemitters2018} in fibers, practical implementations of telecom-wavelength single-photon emitters remain limited\,\cite{ferrentiIdentifyingCandidateHosts2020,zhangMaterialPlatformsDefect2020}. This has been partly attributed to more prominent non-radiative processes in the infrared regime, which reduce the overall emission efficiency\,\cite{turianskyRationalDesignEfficient2024}. These characteristics render visible-range emitters particularly interesting from a scientific perspective. A comprehensive understanding of the origin and energy-level structure of these defects is essential to fully exploit their potential; however, both remain largely unresolved\,\cite{lyonsFirstprinciplesUnderstandingPoint2021a}. Early studies suggested that such defects could arise from crystal dislocations or stacking faults\,\cite{BerhanePhotophysicsofGaN2018}. More recent investigations, however, indicate that the emitters may instead originate from substitutional atoms or complex combinations of substitutional defects\,\cite{Opticaldipolestructure}.

Carbon is one of the most common unintentional dopants in GaN. It is known to introduce deep level defects leading to localized electronic and vibrational levels. Hybrid functional DFT and experimental studies have shown that carbon substituting Nitrogen or Gallium cites ($\text{C}_{\text{N}}$ or $\text{C}_{\text{Ga}}$) can give rise to visible-range luminescence and distinct local vibrational modes\,\cite{lyonsCarbonImpuritiesYellow2010,
mccluskeyLocalVibrationalModes2000,
reshchikovLuminescencePropertiesDefects2005b,gamovCarbonDopingGaN2020}. 

One of the other promising candidates is the $\text{N}_{\text{Ga}}\text{V}_{\text{N}}^0$ complex, which is predicted to exhibit a zero-phonon line (ZPL) in the 600–700~nm range and an excited-state lifetime of approximately 3.5~ns\,\cite{yuanGanAsA2023}, consistent with experimental observations. Nevertheless, further research is required to definitively determine the microscopic nature and energy-level structure of these visible-range defect emitters.

\section{Results and discussion}

\subsection{Linewidth and Auto Correlation}

\begin{figure}[h]
    \centering
    
    \begin{subfigure}{0.48\textwidth}
        \centering
        \includegraphics[width=\linewidth]{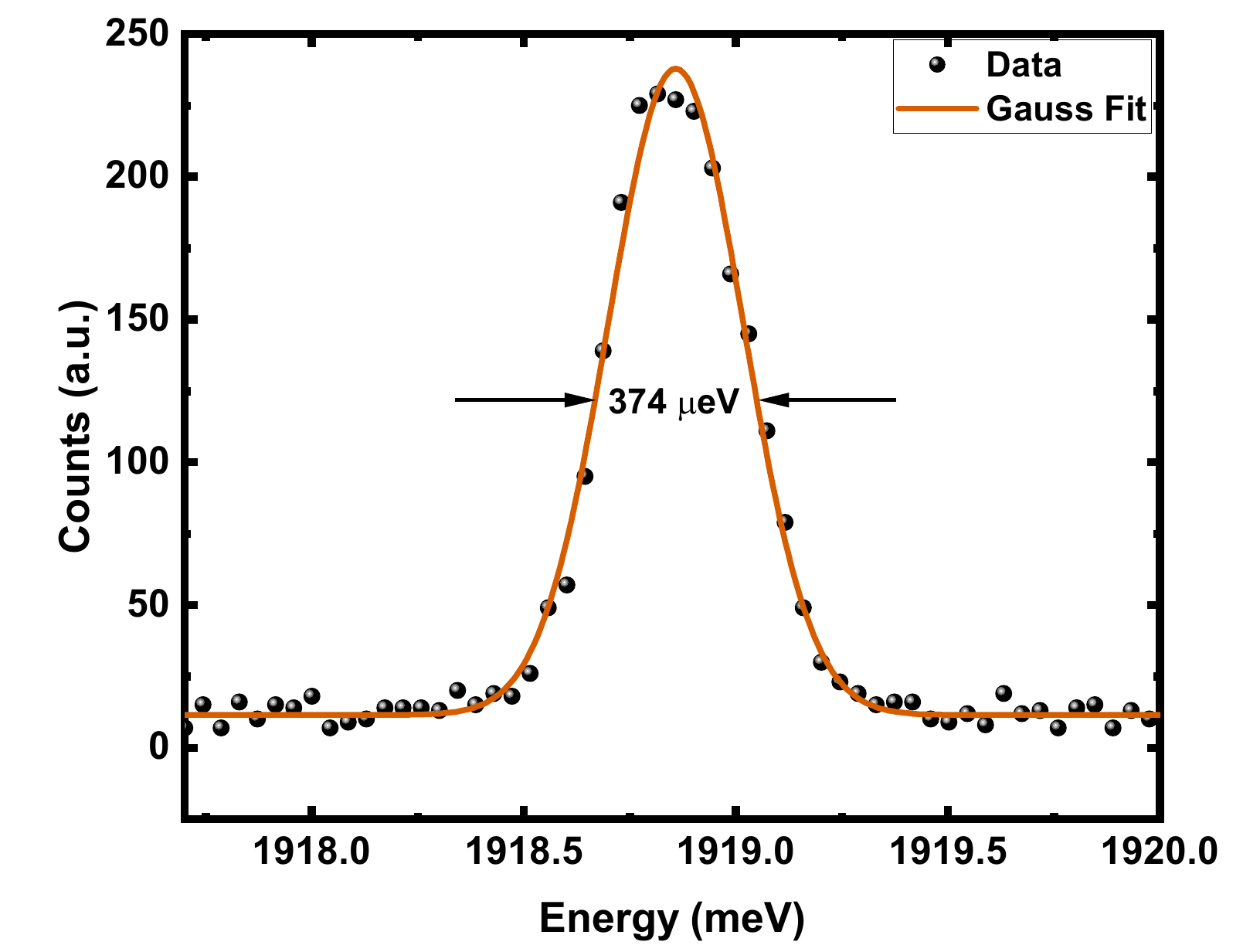}
        \caption{}
        \label{Linewidth}
    \end{subfigure}
    \hfill
    \begin{subfigure}{0.48\textwidth}
        \centering
        \includegraphics[width=\linewidth]{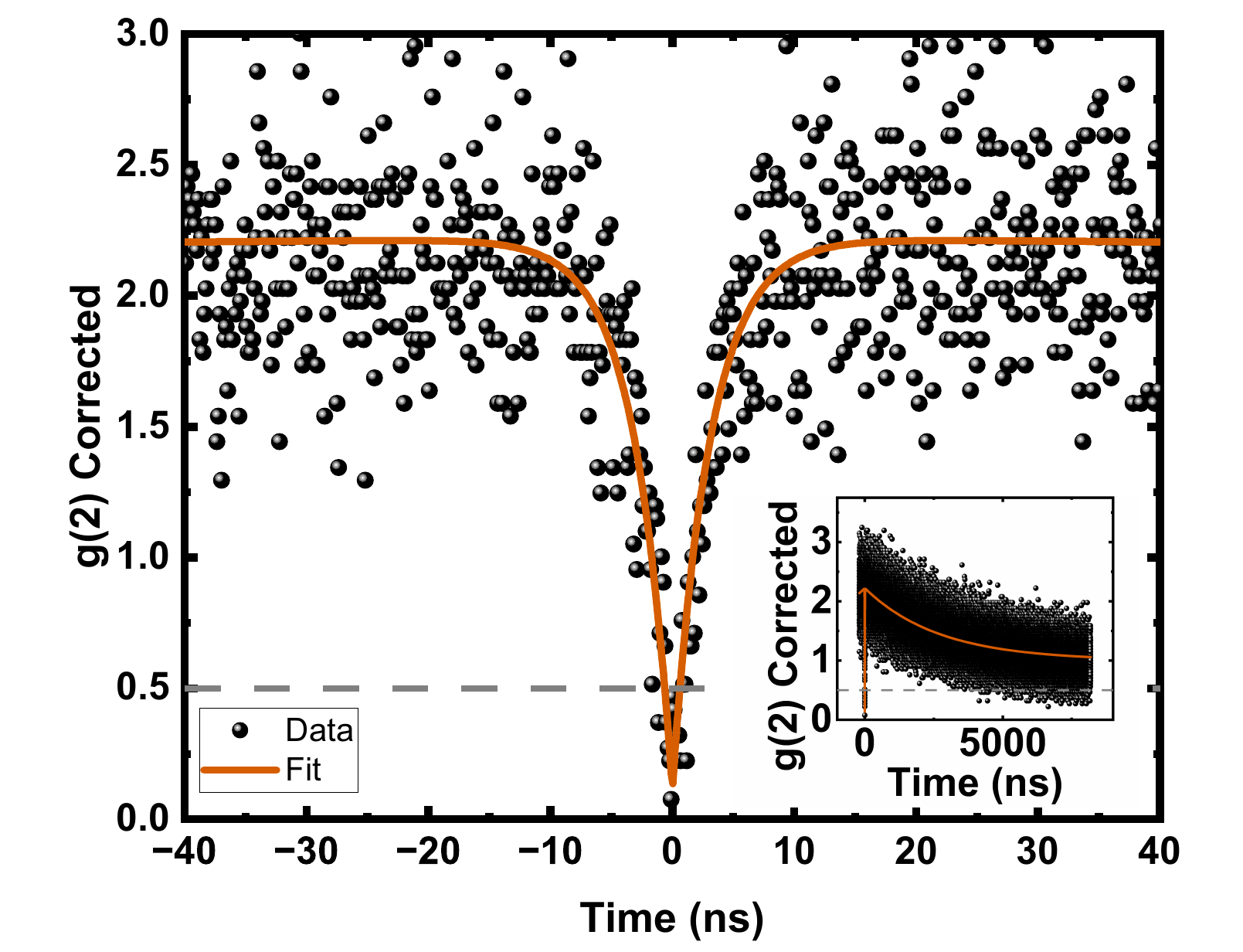}
        \caption{}
        \label{G2}
    \end{subfigure}
    
    \caption{(A) Photoluminescence signal from a single emitter A0. The excitation power was kept at 1\,mW and excitation energy was 2.13 eV (580 nm). Gaussian fit shows a (FWHM) of $374\,\mu$eV. (B) Corrected second-order correlation counts measurement of the same emitter showing in the black color. In the orange color, the fitted curve with $g^{(2)}(0)= 0.11 \pm 0.08$ confirming single-photon emission and fitted values for fast and slow decays as \( t_{\mathrm{fast}} = 3.08 \pm 0.16~\mathrm{ns} \) and \( t_{\mathrm{slow}} = 2698 \pm 11~\mathrm{ns.} \)}The inset shows the long-delay component of the emitter. 
\end{figure}

The sample surface was first characterized at room temperature using a custom-built confocal microscope. Samples exhibiting suitable emitter distributions were subsequently examined at cryogenic temperatures (10 K) by mounting the sample in a cryostat. The emitters were re-identified by scanning the excitation beam across the field of view using a  solid immersion objective\,\cite{jasnyFluorescenceMicroscopySuperfluid1996} or using an aspheric lens mounted on attocube piezo positioners. Quasi-resonant excitation at 2.13 eV (580 nm) was provided by a Rhodamine-based laser, and the emission was spectrally isolated from the excitation using appropriate optical filters before detection with a spectrometer. The excitation power was maintained at 1\,mW. One such exemplary emitter, labelled as A0, is shown in Figure \ref{Linewidth}. The measured full width at half maximum (FWHM) of the emission line was extracted to be 374\,µeV, consistent with values obtained from previously studied samples\,\cite{dallaOffresonantPhotoluminescenceSpectroscopy2025}. In our experiments, we studied also samples with varying carbon concentration however we haven't obtained any conclusive results of impact of carbon concentration on emitter density.

	
	
		
		
		
		
		
	
		
		
		
		

	

 For the same emitter A0, the background corrected counts \( g^{(2)}_{\mathrm{corrected}} \) are shown in Figure\,\ref{G2}. The background-corrected autocorrelation data were fitted using a three-level model (see supplementary information for more information about fit and background correction). The resulting fit, shown in Figure\,\ref{G2} (orange curve), reveals two distinct dynamical components: a fast decay process with characteristic time \( t_{\mathrm{fast}} \) occurring near zero delay, and a slower component with characteristic time \( t_{\mathrm{slow}} \) at longer delays. The pronounced bunching observed around zero delay can be attributed to population trapping in a metastable shelving state. From the fit, we obtain \( t_{\mathrm{fast}}=3.08 \pm 0.16~\mathrm{ns} \) and \( t_{\mathrm{slow}}=2698 \pm 11~\mathrm{ns} \). The extracted value at zero delay is $g^{(2)}(0)= 0.11 \pm 0.08$. Note the error bars here refer to the standard error. The $g^{(2)}(0)$ is well below the classical limit of 0.5, thereby confirming the single-photon nature of the emission.

\subsection{Photoluminescence Excitation (PLE) Scans}

\begin{figure*}[t]
	\centering
	\includegraphics[width=1\linewidth, height=0.5\textheight]{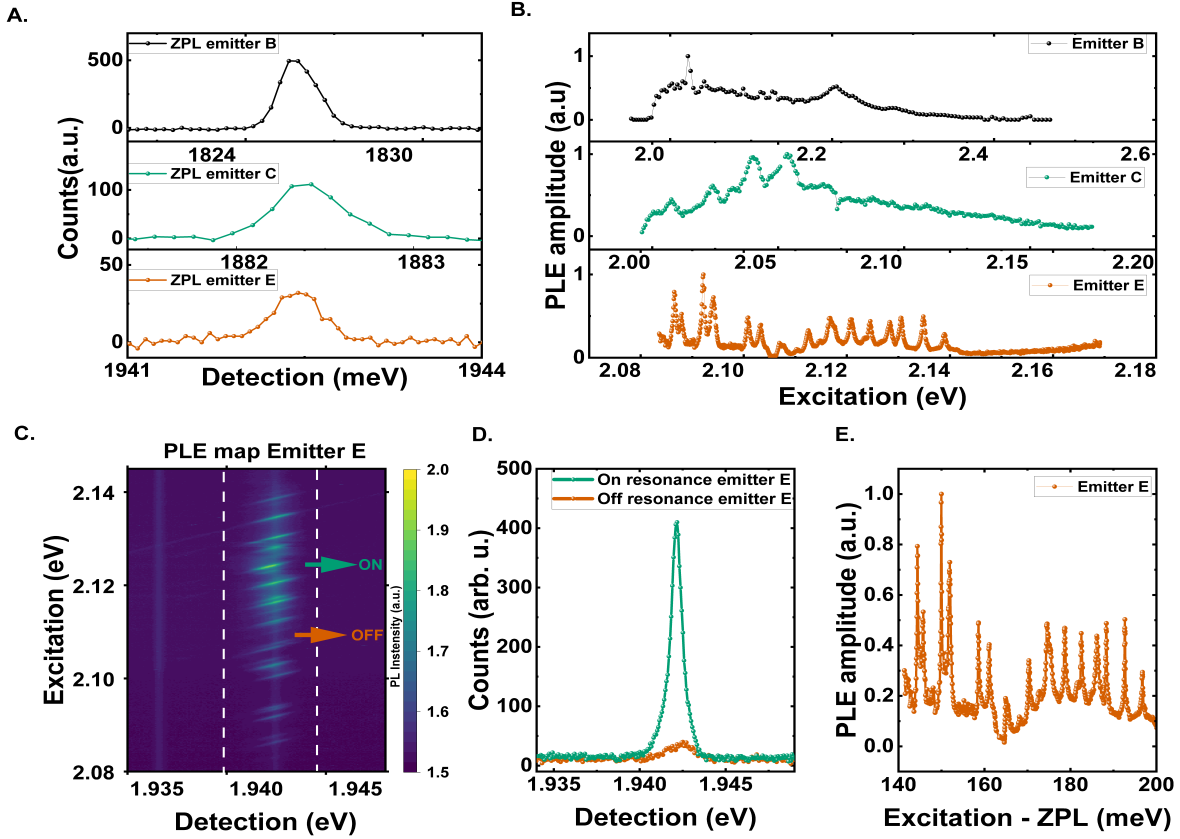}
	\caption{A) ZPL of three shown emitters, namely B, C and E. B) PLE amplitude plotted for the corresponding B, C and E emitters. With emitter E showing strong distinct peaks. C) PLE map of Emitter E with ZPL at 1.942\,eV. On and Off arrows represent cross sections with excitation laser matching or not matching resonance energies. D) On and Off cross section spectra of the PLE map. E) Gaussian fit amplitude of the ZPL (PLE map emitter E) plotted relative to the excitation energy. The distinct peaks correspond to resonance energy values at discrete energy offsets from the ZPL. }
	\label{fig:resonance-1}
\end{figure*}

After confirming our emitters are single-photon emitters (see supplementary information for more correlation results), we conducted a detailed photoluminescence excitation (PLE) study. The excitation energy was scanned over a coarse range from 2.40 to 1.90~eV (520–650~nm) using a supercontinuum source, and over a finer range from 2.17 to 2.10~eV (570–590~nm) using a Rhodamine laser. For each excitation energy, the detection energy range was fixed. The excitation power and sample temperature were maintained at the same values as described earlier. The zero-phonon lines (ZPLs) of three representative emitters are shown in Figure\,\ref{fig:resonance-1}(A), where the labeling of emitters A0, B, C, E and so on follows the convention established in our previous work\,\cite{dallaOffresonantPhotoluminescenceSpectroscopy2025} (the same labels refer to the same emitters in both works). The PLE amplitude was extracted by fitting Gaussian functions to the PLE spectra and is presented in Figure\,\ref{fig:resonance-1}(B). Emitters B and C exhibit a coarse dependence of the excitation efficiency on the excitation energy, with distinct peaks appearing in the 2.00–2.10~eV range. A fine-resolution scan of emitter E within 2.00-2.15~eV spectral window resolved a set of narrow resonances.

For roughly 80 percent of the measured emitters, we observed a pronounced increase in emission intensity as the excitation laser was tuned through specific energies. This enhancement is evident in the PLE map of emitter~E, presented in Figure\,\ref{fig:resonance-1}(C). When the excitation energy is resonant with discrete excited states of the emitter, sharp and narrow spectral features emerge, intersecting the zero-phonon line (ZPL). Such narrow resonances are not universal across all emitters and, in selected cases, appear significantly weaker. Representative cross-sections of the PLE map, corresponding to on-resonance and off-resonance excitation (marked ``On'' and ``Off'', respectively), are shown in Figure\,\ref{fig:resonance-1}(D). Excitation via the strongest resonance results in about an order of magnitude enhancement of the ZPL emission.  

To quantify the detuning of the narrow resonances from the ZPL, a subsection of the PLE map (indicated between white dashed lines in Figure\,\ref{fig:resonance-1}(C) was analyzed and PLE amplitude fit is shown in Figure\,\ref{fig:resonance-1}(E). The data reveals that resonances occur at discrete energy offsets from the ZPL. Furthermore, the brightness of these resonances varies, with some exhibiting significantly higher intensity than others. Finally, the linewidth of the resonances was measured to be approximately 0.5~meV. In the limiting case of purely homogeneous (lifetime-limited) broadening, this would correspond to a lifetime on the order of 1.3~ps

\subsection{Polarization-resolved PLE study}

\begin{figure*}[t]
	\centering
	\includegraphics[width=1\linewidth, height=0.4\textheight]{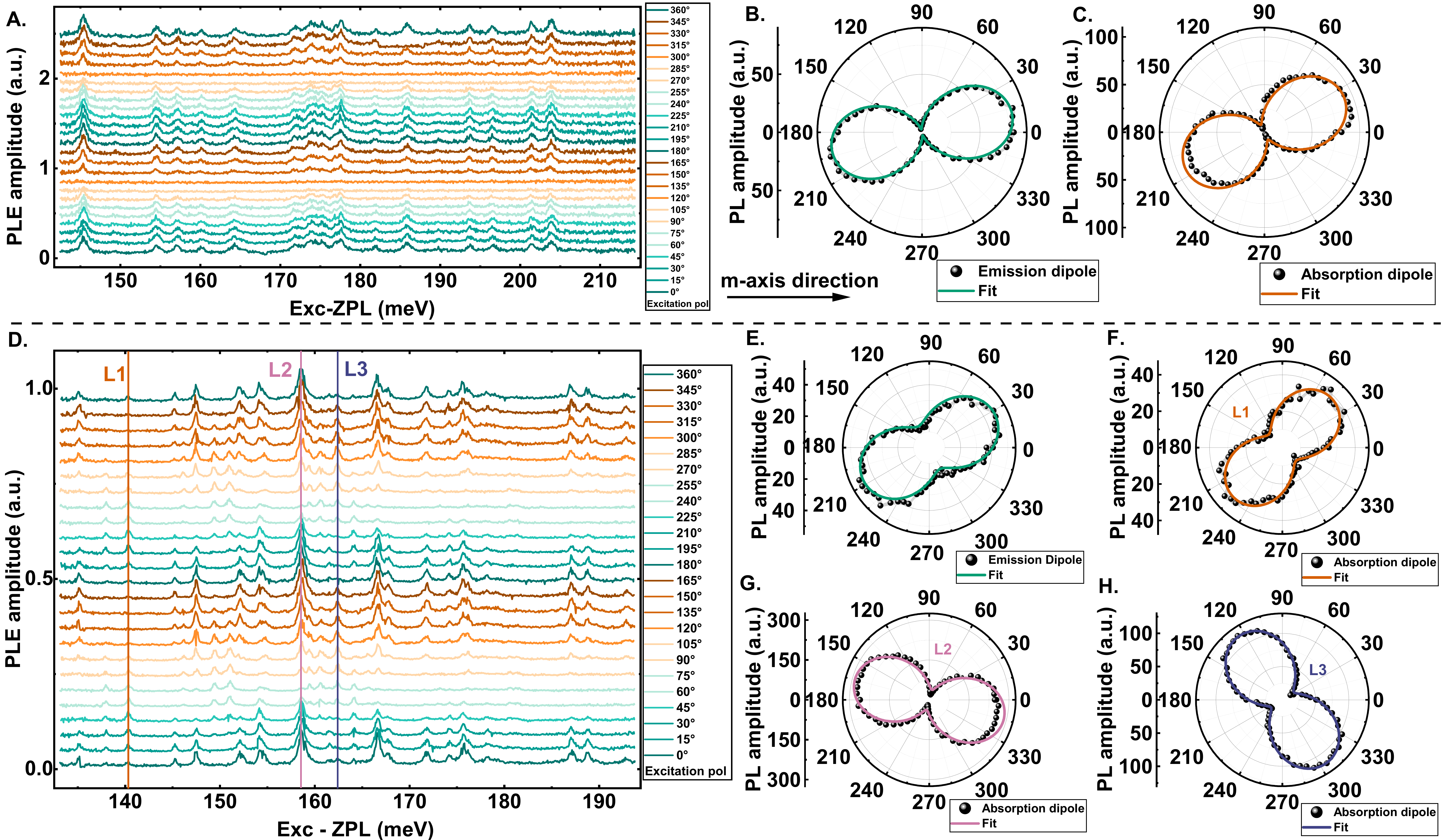}
	\caption{A) Angle resolved PLE spectra of emitter F, with resonance energies shown with respect to the ZPL. All shown angles are with respect to the GaN in plane m axis direction. B) Emission dipole pattern plotted for the emitter F. C) Absorption emission pattern of emitter F. D) Angle resolved PLE spectra of another emitter G shown with resonance energies shown with respect to the ZPL. Lines L1, L2 and L3 denote three different resonance energies. E) Emission dipole pattern of emitter G. \\
    F-H.) Absorption dipole pattern measured at three different resonance energies namely at L1, L2 and L3 shown in (D).}
	\label{fig:resonance-2}
\end{figure*}

To gain further insight into the nature of the observed resonances, we performed polarization-resolved PLE measurements. For excitation polarization control, the laser beam was passed through a linear polarizer followed by a half-wave (\(\lambda/2\)) plate. Then for each excitation angle the excitation energy was swept. Similarly, to characterize the emission polarization, a half-wave plate and a linear polarizer were placed in the detection path before the spectrometer. In order to record angle resolved PLE plots for each emitter, the emission signal was collected through a monochromator on an avalanche photodiode unit. All the angles measured as shown in Figure\,\ref{fig:resonance-2} are measured from the GaN in-plane m-axis. All the collected spectra have been power-normalized since we operate in a linear power dependence regime. Then onward, for a given excitation energy, the excitation/detection polarization was rotated and resulting spectrum was recorded on the spectrometer. The recorded spectrum was then fitted with Gaussian peak and the extracted amplitude was recorded and plotted to show the dipole pattern of the emitter in absorption or detection.  All the dipoles were then fitted with $A+B.cos^2 (\phi + \phi_0)$ to deduce the dipole angle ($\phi_0$). In this experiment, we studied 44 different emitters in this manner and we find two such group of emitters explained below.

Angle resolved PLE spectra of an emitter F has been shown in the Figure\,\ref{fig:resonance-2}\,(A). It can be seen that a number of peaks appear in the PLE spectra of the emitter when the excitation is kept at 0 degrees. As excitation polarization direction is tuned all the resonances disappear together at about 120 degrees. From this point on-wards, all of the resonances appear together and reach a maximum at about 195 degrees. Therefore, for this emitter, all the resonances are polarized in the same direction. Figure\,\ref{fig:resonance-2}\,(B) shows the recorded emission pattern and Figure\,\ref{fig:resonance-2}\,(C) shows the absorption dipole pattern of emitter F. The absorption pattern lies in agreement with the full angle resolved PLE plot. This is the most commonly observed behavior of the emitters that we studied. Such emitters that have the resonances polarized in the same direction, we classify as group 1 emitters.

Another example of a different emitter, Emitter G, is shown in Figure\,\ref{fig:resonance-2}\,(D-H). The polarization-resolved PLE is shown in Figure\,\ref{fig:resonance-2}\,(D). At first glance, many resonance peaks are present for this given emitter. However on a careful inspection it shows that there are three different group of peaks in this plot. Each peak on this plot falls into one of the three groups. Each peak group behaves slightly differently when the polarization direction is changed. One such line of each group is represented by L1, L2 and L3 on the plot. The corresponding absorption dipole patterns for each line L1-L3 are plotted separately in Figure\,\ref{fig:resonance-2}\,(F-H) and marked with same colors correspondingly. We observe that there are 3 different absorption dipole orientations that this emitter can have depending on the excitation energy. Emission dipole pattern of the emitter has been also plotted in Figure\,\ref{fig:resonance-2}\,(E). The emission dipole remains the same regardless of the excitation energy. Such emitters are classified as group 2 emitters, where the absorption dipole orientation is no longer unique and depends on the excitation energy.
We do not observe any correlation  between emitter group (1 or 2) and excitation energy, emission energy, or emission dipole orientation relative to the GaN high-symmetry axes.  

\begin{figure*}[t]
	\centering
	\includegraphics[width=1\linewidth]{"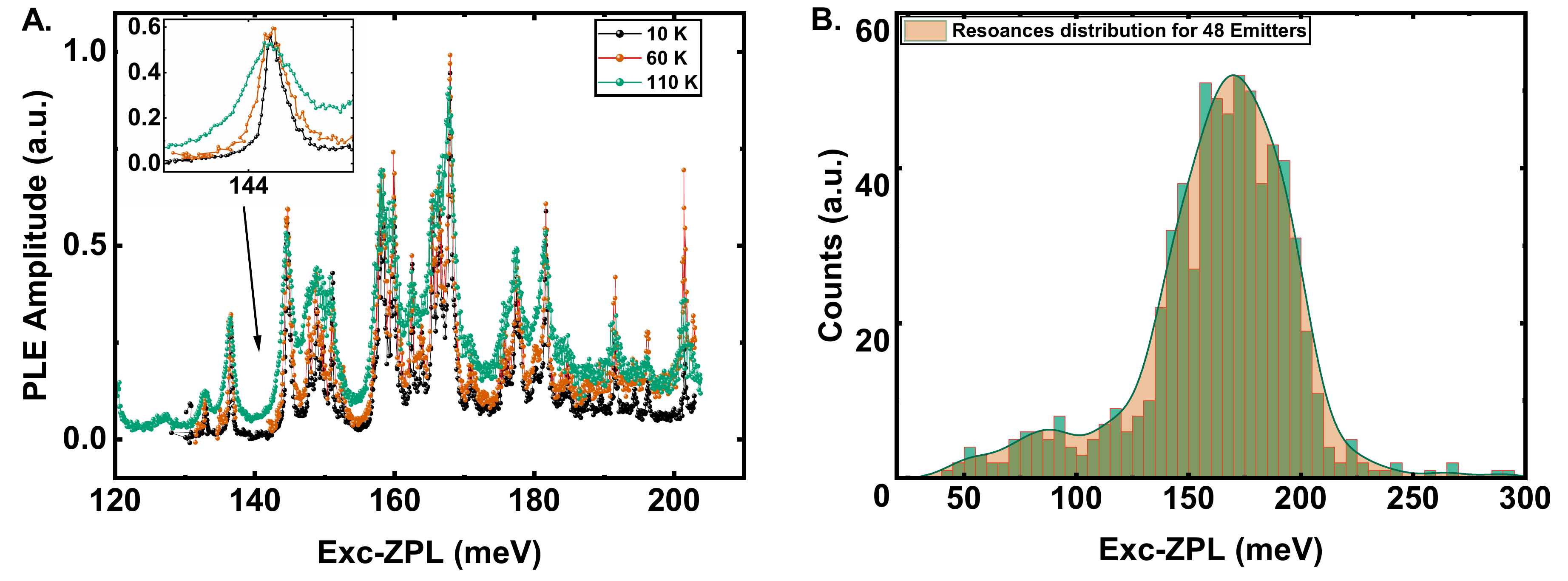"}
	\caption{A.) Resonance energies shown for an emitter H at different temperatures namely 10, 60 and 110 K. The inset shows a zoom at around 144 meV energy difference. B.) Distribution of resonance peak energies calculated as difference of excitation energy and ZPL energy for 48 different emitters. } 
	\label{fig:resonance-3}
\end{figure*}

\subsection{Temperature and excitation energy dependence of the resonances}

To investigate the temperature dependence of the resonances, PLE scans were repeated at three different temperatures for emitter H. The extracted resonance positions relative to the excitation energy are presented in Figure\,\ref{fig:resonance-3}\,(A), with the inset providing a magnified view of the spectral region near 144~meV. The data show that the resonance positions remain unchanged with temperature. However, the overall emission intensity decreases with increasing temperature, accompanied by a slight broadening of the resonance peaks. This behavior may indicate a phonon-mediated contribution to the observed resonances.

To determine whether the observed resonances are specific to individual emitters, we performed a systematic study across multiple emitters. Not all emitters exhibited resonances, and among those that did, the resonance energies varied between emitters. In this experiment, we show the data of 48 different emitters displaying statistically significant resonance peaks. The energies of the resonance peaks were plotted on a relative energy scale, and a histogram was constructed with a bin width of 5~meV, as shown in Figure\,\ref{fig:resonance-3}\,(B). Resonances were observed within the 50–300~meV excitation range relative to the ZPL position, with an apparent higher density of states around 170~meV. So far, in our measurements, no correlation was observed between the ZPL energies of the emitters and the presence or energies of the resonances in our measurements.

Furthermore, we have investigated multiple emitters across different samples and, to date, have not observed any irreversible modification or disappearance (“bleaching”) of the resonance patterns. This indicates that the resonances are intrinsically linked to the emitter and persist as long as the emitter remains optically active. 

\begin{figure*}[t]
	\centering
	\includegraphics[width=1\linewidth]{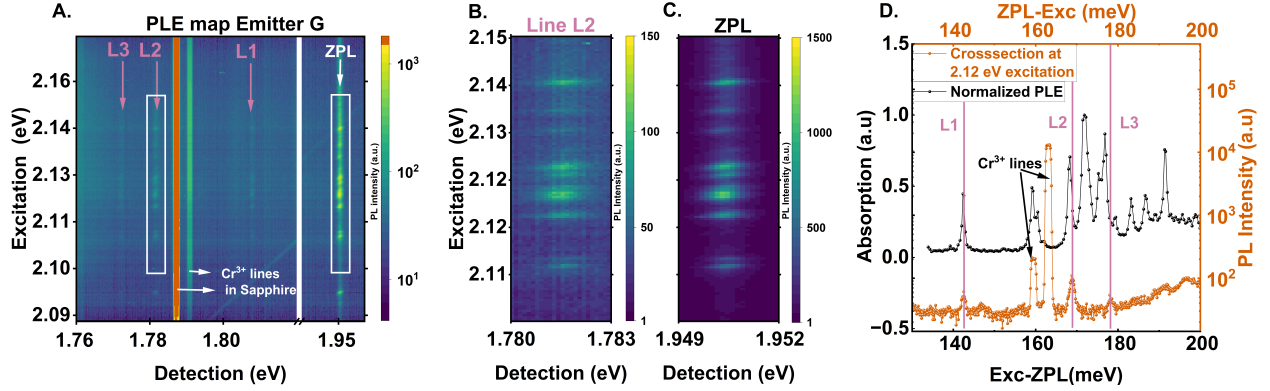}
	\caption{A.) PLE map of a yet another emitter I showed in the left panel. The ZPL and $Cr^{3+}$ lines from the sapphire has been marked accordingly (orange color marks the saturation region). The lines L1, L2 and L3 are marked in the PLE map, which correspond to low energy peaks. When the resonances transition through ZPL, faint L1, L2 and L3 lines are also seen to get brighter. B)-C) Zoomed in view of the lines L2 and ZPL (marked by white	rectangle on map) have been presented. D) The rightmost panel shows the PLE amplitude	of the ZPL in the black color with respect to the ZPL energy. The cross-section of the PLE map is also plotted in the orange color. As can be seen the lines L1, L2, L3 are mirrored at the same energy distance on the lower energy side of the ZPL. } 
	\label{fig:resonance-4}
\end{figure*}

\subsection{Low energy peaks}

In order to examine whether we see the same absorption lines in the photoluminescence signal as well we moved our detection window towards lower energies during the PLE scans. One such exemplary PLE map, for emitter I, is shown in Figure\,\ref{fig:resonance-4}\,(A). The map highlights a few important features. The ZPL as marked is located at about 1.945\,eV. The excitation energy is swept from 2.09 to 2.18\,eV. It can be seen the ZPL passes through several resonances in this scanning range. Also on the lower detection energy, three faint lines marked as L1, L2 and L3 can be seen. Their position mirrors that of the absorption peaks. The lines are very dim in nature but the intensity of these lines is positively correlated with the intensity of the ZPL. The $Cr^{3+}$ lines in sapphire are also marked at detection energy of about 1.769~eV. During all PLE measurements, the excitation power was monitored and the extracted PLE amplitudes were normalized by the measured laser power. Power-dependent measurements confirmed that the emitters were operated in the linear excitation regime, while the faint background fluctuations visible in the PLE maps may arise from background channels that may not be in the same linear regime. The Figure\,\ref{fig:resonance-4}\,(B,C), shows the zoomed-in view of the lower energy line L2 and ZPL of the emitter. The zoomed-in area is marked with a white rectangle in Figure\,\ref{fig:resonance-4}\,(A). As also evident from here the intensity pattern of the line and ZPL is strongly correlated.

The rightmost graph Figure\,\ref{fig:resonance-4}\,(D), shows the PLE amplitude of the ZPL in the black color. As can be seen the resonances are present in the range of 140-200~meV away from the ZPL position. The orange line shows a cross-section through the PLE map at 2.12~eV excitation energy plotted as the PL emission. The $Cr^{3+}$ lines from sapphire substrate are also visible in the spectra. The PL emission lines occur at energies corresponding to the resonances observed in absorption, although a perfect spectral overlap is not observed. Additionally, we note that not all absorption resonances have corresponding features in the PL emission.

\subsection{Discussion}
Although the complete microscopic origin of the resonances is not yet fully established, our results offer relevant clues and enable a discussion of their possible origins.

Bulk phonons are one of the candidates to explain the observed resonances. However, their energies in GaN occur at significantly lower energy scales (approximately 70~meV for the $E_2$ (high) mode and 92~meV for the $E_1$(LO) mode), indicating that the observed resonances cannot be attributed to coupling with the principal bulk phonons. The 140–200~meV energy range corresponds well with the multi-phonon absorption\,\cite{gamovFingerprintsCarbonDefects2022}. However, the relative resonance energies varies between emitters and no universal pattern could be attributed to a combination of two well-defined (at $\Gamma$-point) phonons (see Figure\,\ref{fig:resonance-2}~F). 

We hypothesize that the resonances may be associated with localized vibrational modes (LVMs). The defects responsible for single-photon emission in the visible range in GaN are typically complexes of substitutional impurities and vacancies\,\cite{Opticaldipolestructure,yuanGanAsA2023}. In particular, theoretical calculations for the $\text{N}_{\text{Ga}}\text{V}_{\text{N}}^0$ defect complex suggest that localized vibrational levels may exist at energies corresponding to those of the observed resonances\,\cite{yuanGanAsA2023}. The large energy of the LVM would indicate coupling to a light-atom–containing defect. We note that multiple peaks can not be overtones of one LVM, which is in agreement with the small Huang-Rhys factor observed for these defects\,\cite{berhaneBrightRoomTemperature2017}. A plausible explanation would be a presence of multiple local modes, also supported by their polarization selectivity that could be associated with the geometry of the defect. The measured linewidth of the resonances (0.5~meV)  and the occasional presence of the emission lines on the lower energy side of ZPL (mirroring the absorption peaks, as shown in Figure\,\ref{fig:resonance-3}) further support the hypothesis of associating the observed resonances to LVM. 

 Several studies indicate that donor-acceptor pairs (DAP) are responsible for single-photon emission in hexagonal Boron Nitride (hBN). One study explicitly states that most single-photon emitters in hBN can be well explained by donor-acceptor pairs, with their wavelength fingerprints matching the experimentally observed photoluminescence spectrum\,\cite{tanDonorAcceptorPair2022a}. Potentially, a similar mechanism could be responsible for the multiplicity of the emission energy of the GaN single-photon defect centers. However,  the donor-acceptor pair band transitions in GaN are situated at around 3~eV energy range\,\cite{reshchikovLuminescencePropertiesDefects2005a}, so well above the energy of the ZPL observed in our experiments. 
 With rising temperature, the ZPL observed in our experiments monotonically redshift by around 3 meV (from 4 to 180 K), closely resembling the behavior described in Ref\,\cite{gengDephasingOpticalPhonons2023}.
The energy difference between the observed resonances and ZPL is constant with temperature, as observed in Figure\,\ref{fig:resonance-3} (A). The change of energy of the donor-acceptor pair with temperature can be non-monotonic\,\cite{ yuBandGapPhotoreflectance2004} and shift in even up to 100 meV-scale\,\cite{paskovaDonoracceptorPairEmission2005}. Such observations point toward associating the observed resonances with LVM, rather than phenomena related to DAP.  

The broad distribution of resonance energies, rather than a uniform pattern across emitters, may be linked to variations in sample quality. Carbon doping can introduce multiple imperfections, which most likely contributes to the higher emitter density observed in our samples. Consequently, individual emitters may experience slightly different local environments, resulting in variations in resonance energies. A complete understanding of the origin of these resonances and their relationship to specific defect configurations will require further theoretical investigation. We hope that our experimental evidences will allow to restrict the theoretical search for specific defect centers.

\section{Conclusions}
In this work, we report experimental observation of resonances in photoluminescence excitation of GaN single-photon emitters operating in the visible spectral range. The resonances are likely related to phonon-assisted excitation. We demonstrate that quasi resonant excitation through these states can enhance the zero-phonon line (ZPL) emission of individual emitters by up to an order of magnitude. The resonances can be selectively addressed via the polarization of the excitation light, while the emission dipole orientation of the emitter remains unchanged. Statistical analysis indicates that these resonances occur most frequently around 170~meV. We attribute the origin of the resonances to coupling between the emitter and localized phonon modes, which gives rise to sharp spectral features in the photoluminescence excitation (PLE) spectra. Consequently, these resonant states provide a quasi-resonant excitation pathway, enabling increased emission into the ZPL. Our findings offer a pathway to enhance the brightness of GaN single-photon emitters and provide valuable insight into the energy-level structure of the underlying defects, facilitating their integration into photonic platforms.

\section{Experimental Setup}

The GaN sample studied in this work consists of two GaN layers ($1.5\,\mu m$ buffer layer and $3.5\,\mu m$ of controlled carbon concentration layer) grown on a sapphire substrate using an Aixtron CCS-FT 3x2 MOCVD system. The carbon concentration in the sample was $2.3\times10^{17}/cm^3$. To estimate the emitter density, we used a room-temperature confocal scanning microscopy setup where the sample was illuminated with a 532 nm green laser diode and the signal was collected on an avalanche photo-diode after spectral filtering of the detected signal. To confirm the single photon nature of the emitters we used two PerkinElmer fast photo diodes mounted at the exit of two 700 mm long spectrometers in Czerny-Turner configuration, equipped with gratings up to 2400 lines/mm and a CCD camera featuring pixel size of $16\,\mu m$. The excitation photon energy was continuously monitored using a calibrated wavemeter, providing high-precision determination of the excitation energy. The emission energies, including the zero-phonon line (ZPL) and lower-energy spectral features, were measured using the spectrometer that was calibrated using known reference lines prior to the measurements.

For correlation measurements, emission signal from selected emitters was equally divided using a 50:50 non-polarizing beam splitter and directed to two PerkinElmer single-photon avalanche photodiodes (SPADs) positioned at the output ports of successive stages of the \textit{TriVista} spectrometer. Second-order photon autocorrelation measurements were performed using a PicoQuant \textit{HydraHarp} time-correlated single-photon counting module.

The high refractive index of GaN decreases the setup's collection efficiency. For experiments requiring high number of photon counts we employed a cryo-compatible immersion objective\,\cite{jasnyFluorescenceMicroscopySuperfluid1996} in contact with the GaN sample surface to address this issue. This approach also allowed us to significantly minimize chromatic aberrations during photoluminescence excitation studies. The sample and excitation objective were mounted inside an Oxford Spectromag helium-bath cryostat. For broadband excitation spectroscopy, a supercontinuum source was employed to cover the energy range from 2.40 to 1.90 eV (520–650 nm). For high-resolution excitation scans, a tunable dye laser was used, providing continuous coverage from 2.03 to 2.17 eV (570–610 nm). The detection channel was spectrally filtered using interference filters to isolate emission in the range of 1.77–2.06 eV (600–700 nm). A linear polarizer and a rotatable half-wave plate were incorporated into both the excitation and detection optical paths to enable complete characterization of the emitters’ absorption and emission polarization properties.

\medskip
\textbf{Acknowledgements} \par 
Authors thank Marek Maciaszek and Ivan Gamov for insightful discussions. 
This work was partially supported by the Polish National Science Center (NCN) under Decision DEC-2021/43/D/ST7/03367. N.D., P.K., M.K. and T.J, acknowledge support from the Polish National Agency for Academic Exchange (NAWA) under Polish Returns 2019 Programme (Grant No. PPN/PPO/2019/1/00045/U/0001). TJ acknowledges funding from the European Horizon EIC Pathfinder Open programme under grant agreement no. 101130384 (QUONDENSATE).

\medskip
\textbf{Conflict of Interest} \par
The authors declare no conflict of interest.

\medskip
\textbf{Supporting information} \par
Supporting information document related to the article can be found in the attachments.

\medskip
\textbf{Data Availability Statement} \par
The data that support the findings of this study are available from the corresponding author upon reasonable request.

\medskip

%


\bibliography{references}

\end{document}